\documentclass[runningheads]{svjour2}
\smartqed  
\usepackage{graphicx}
\journalname{General Relativity and Gravitation}

\begin{document}

\title{Anisotropic massive strings in scalar-tensor theory of gravitation}

\author{Anil Kumar Yadav}
\authorrunning{Yadav A. K.}

\institute{Anil Kumar Yadav \at Department of Physics, Anand
Engineering College, Keetham, Agra 282 007,
India\\\email{abanilyadav@yahoo.co.in}}

\date{Accepted . Received ; in original form }

\maketitle

\begin{abstract}
We present the model of anisotropic universe with string fluid as source of matter 
within the framework of scalar-tensor theory of gravitation. Exact solution of 
field equations are obtained by applying Berman's law of variation of Hubble's parameter 
which yields a constant value of DP. The nature of classical potential is examined for the 
model under consideration. It has been also 
found that the massive strings dominate 
in early universe and at long last disappear from universe. This is in agreement 
with current astronomical observations. The physical and dynamical properties of model are also discussed.\\ 
\end{abstract}

\section{Introduction}
On the basis of coupling between an adequate tensor field and scalar field $\phi$, 
Brans and Dike\cite{Brans1961} formulated the scalar-tensor theories of gravitation.
The scalar field $\phi$ has the dimension of $G^{-1}$ therefore $\phi^{-1}$ 
plays the role of time varying gravitational constant $G$. 
This theory is more consistent with Mach's principle and less 
reliant on the absolute properties of space. A detail survey of Brans-Dike theory has been 
done by Singh and Rai\cite{Singh1983}. In fact the notion of time-dependent $G$ was first 
conceived by Dirac\cite{Dirac1938}, though Dirac's arguments were based on cosmological 
considerations not directly concerned with Mach's principle.\\

In Brans-Dike theory, which is the generalisation of general relativity, 
an additional scalar field $\phi$ besides the metric tensor $g_{ij}$ and 
a dimensionless coupling constant $\omega$ were introduced. For large value of 
coupling constant $\omega$ (i. e. $\omega > 500$), Brans-Dike theory follows the result of 
general relativity. 
Later on, Saez and Ballester\cite{Saez1985} developed a scalar-tensor theory in 
which a dimensionless scalar field is coupled with metric. 
This coupling use to give a satisfactory description of the weak fields. 
This scalar-tensor theory play an important role to solve the missing matter 
problem and to remove the graceful exit problem in non flat FRW cosmologies 
and inflation era\cite{Piemental1997} respectively. 
The following authors, Singh and Agarwal\cite{Singh1991,Singh1992}, Reddy et al\cite{Reddy2006,Reddy2008}, 
Socorro and Sabibo\cite{Socorro2010} and recently Jamil et al\cite{Jamil2012} have studied cosmological model 
within the framework of Sa'ez-Ballester scalar-tensor theory of gravitation in different physical contexts.\\ 

Among the different cosmological structure of universe, the cosmic string models 
have wide acceptance because it give rise to density perturbations which lead to 
formation of galaxies\cite{Vilenkin1985}. Firstly, Letelier\cite{Latelier1979} described 
the gravitational effect of massive strings which are formed by geometric strings with particles 
attached along their extension. At the observational front, Pogasian et al\cite{Pogosian2003} have 
showed that the cosmic strings are not responsible for either the CMB fluctuations or the observed 
clustering of galaxies. Recently Yadav et al\cite{Yadav2011} and Yadav\cite{Yadav2012} have studied 
Bianchi-V string cosmological models in general relativity. 
In this paper, we discuss Einstein's field equations in scalar tensor theory of gravitation 
for Bianchi - V space-time, filled with string fluid as source of matter. Exact solution of 
field equations are obtained by applying the law of variation of Hubble's parameter, firstly 
proposed by Berman\cite{Berman1983}. This law yield the constant value of DP.

\section{Matric and Basic equations}
The spatially homogeneous and anisotropic Bianchi-V space-time 
is described by the line element
\begin{equation}
 \label{eq1}
ds^{2}=-dt^{2}+A^{2}dx^{2}+e^{2\alpha x}\left(B^{2}dy^{2}+C^{2}dz^{2}\right)
\end{equation}
where $A(t)$, $B(t)$ and $C(t)$ are the scale factors in different spatial directions and 
$\alpha$ is a constant.\\
We define the average scale factor $(a)$ of Bianchi-type V model as
\begin{equation}
 \label{eq2}
a=(ABC)^{\frac{1}{3}}
\end{equation}
The spatial volume is given by
\begin{equation}
\label{eq3}
V = a^{3} = ABC 
\end{equation}
Therefore, the mean Hubble's parameter $(H)$ read as 
\begin{equation}
 \label{eq4}
H=\frac{\dot{a}}{a}=\frac{1}{3}\left(H_{1}+H_{2}+H_{3}\right)
\end{equation}
where $H_{1}=\frac{\dot{A}}{A}$, $H_{2}=\frac{\dot{B}}{B}$ and $H_{3}=\frac{\dot{C}}{C}$ are the 
directional Hubble's parameters in the direction of $x$, $y$ and $z$ respectively. An over dot denotes 
differentiation with respect to cosmic time t.\\

We define the kinematical quantities such as expansion scalar $(\theta)$, 
shear scalar $(\sigma)$ and anisotropy parameter $(A_{m})$ as follows:
\begin{equation}
 \label{eq5}
\theta = u;^{i}_{i}
\end{equation}
\begin{equation}
 \label{eq6}
\sigma^{2} = \frac{1}{2}\sigma_{ij}\sigma^{ij}
\end{equation}
\begin{equation}
 \label{eq7}
A_{m} = \frac{1}{3}\sum_{i =1}^{3}\left(\frac{H_{i}-H}{H}\right)^{2}
\end{equation}
where $u^{i} = (0,0,0,1)$ is a matter four velocity vector and 
\begin{equation}
 \label{eq8}
\sigma_{ij} = \frac{1}{2}\left(u_{i; \alpha}P^{\alpha}+u_{j;\alpha}P^{\alpha}_{i}\right)-\frac{1}{3}\theta P_{ij}
\end{equation}
Here, the projection vector $P_{ij}$ has the form
\begin{equation}
\label{eq9}
P_{ij} = g_{ij} - u_{i}u_{j}
\end{equation}
The expansion scalar $(\theta)$ and shear scalar $(\sigma)$, in Bianchi-V space-time, have the form
\begin{equation}
 \label{eq10}
\theta = 3H = \frac{\dot{A}}{A}+\frac{\dot{B}}{B}+\frac{\dot{C}}{C}
\end{equation}
\begin{equation}
 \label{eq11}
2\sigma^{2} = \left[\left(\frac{\dot{A}}{A}\right)^{2}+\left(\frac{\dot{B}}{B}\right)^{2}+
\left(\frac{C}{C}\right)^{2}\right]-\frac{\theta^{2}}{3}
\end{equation}
Here, $(;)$ stands for covariant derivative with respect to cosmic time $t$.\\
\section{Field equations}
We consider homogeneous and anisotropic Bianchi-V metric coupled with 
scalar field $\phi$. Our model is based on Saez-Ballester theory of gravitation 
which is based on coupling of dimensionless scalar field with metric.\\ 

We assume the Lagrangian
\begin{equation}
 \label{eq12}
L = R-\omega\phi^{r}\phi_{,i}\phi^{,i}
\end{equation}
where $R$, $\omega$ and $r$ represent the scalar curvature, coupling constant and 
dimensionless arbitrary constant respectively.\\
For the scalar field having the dimensions of $G^{-1}$, the Lagrangian (\ref{eq12}) is 
not physically admissible because two terms of the right hand side of equation (\ref{eq12}) 
have different dimension. However, it is suitable Lagrangian in the case of 
dimensionless scalar field.\\
From the above Lagrangian, we can establish the action
\begin{equation}
 \label{eq13}
I = \int_{\Sigma}(L+8\pi L_{m})(-g)^{\frac{1}{2}}dx^{1}dx^{2}dx^{3}dx^{4}
\end{equation}
where $L_{m}$ is the matter Lagrangian, $g$ is the determinant of the matrix $g_{ij}$, 
$x^{i}$ are the coordinates and $\Sigma$ is an arbitrary region of integration.\\
The variational principle $\delta I = 0$ 
leads to the field equations
\[
 G_{ij}-\omega \phi^{r}\left(\phi_{,i}\phi_{,j}-\frac{1}{2}g_{ij}\phi_{,\ell}\phi^{,\phi}\right) = -8\pi T_{ij},
\]
\begin{equation}
 \label{eq14}
2\phi^{r}\phi^{,i}_{;i}+r\phi^{r-1}\phi_{,\ell}\phi^{,\ell}=0
\end{equation}
Equation (\ref{eq14}) is obtained by considering arbitrary independent variations of the metric and scalar field 
vanishing at the boundary of $\Sigma$.\\

Since, the action $I$ is a scalar, it can easily proved that the equation of motion 
\begin{equation}
 \label{eq15}
T^{ij}_{;i} = 0
\end{equation}
are consequences of the field equations.\\
The energy momentum tensor for a cloud of massive strings 
and perfect fluid distribution is taken as
\begin{equation}
 \label{eq16}
T_{ij}=(\rho+p)u_{i}u_{j}+pg_{ij}-\lambda x_{i}x_{j}
\end{equation}
where $p$ is isotropic pressure; $\rho$ is the proper energy density for 
the cloud of strings with particle attached to them; $\lambda$ is the string 
tension density; $x^{i}$ is a unit space-like vector representing the direction 
of string.\\

Choosing $x^{i}$ parallel to $\frac{\partial}{\partial x}$, we have
\begin{equation}
 \label{eq17}
x^{i} = (A^{-1},0,0,0)
\end{equation}
Here, the cosmic string has been directed along x-axis.\\

If the particle density of the configuration is denoted by 
$\rho_{p}$, then
\begin{equation}
 \label{eq18}
\rho = \rho_{p}+\lambda
\end{equation}
The Einstein's field equations (in gravitational units $c = 1$, $8\pi G = 1$)
\begin{equation}
 \label{eq19}
R_{ij}-\frac{1}{2}R g_{ij} = -T_{ij}
\end{equation}
The Einstein's field equations (\ref{eq19}) for the line-element 
(\ref{eq1}) lead to the following system of equations
\begin{equation}
 \label{eq20}
\frac{\ddot{B}}{B}+\frac{\ddot{C}}{C}+\frac{\dot{B}\dot{C}}{BC}-\frac{\alpha^{2}}{A^{2}} = 
-p+\frac{1}{2}\omega\phi^{r}{\dot{\phi}}^{2}+\lambda
\end{equation}
\begin{equation}
 \label{eq21}
\frac{\ddot{A}}{A}+\frac{\ddot{C}}{C}+\frac{\dot{A}\dot{C}}{AC}-\frac{\alpha^{2}}{A^{2}} = 
-p+\frac{1}{2}\omega\phi^{r}{\dot{\phi}}^{2}
\end{equation}
\begin{equation}
 \label{eq22}
\frac{\ddot{A}}{A}+\frac{\ddot{B}}{B}+\frac{\dot{A}\dot{B}}{AB}-\frac{\alpha^{2}}{A^{2}} = 
-p+\frac{1}{2}\omega\phi^{r}{\dot{\phi}}^{2}
\end{equation}
\begin{equation}
 \label{eq23}
\frac{\dot{A}\dot{B}}{AB}+\frac{\dot{A}\dot{C}}{AC}+\frac{\dot{B}\dot{C}}{BC}-\frac{3\alpha^{2}}{A^{2}} = 
\rho-\frac{1}{2}\omega\phi^{r}{\dot{\phi}}^{2}
\end{equation}
\begin{equation}
 \label{eq24}
\frac{2\dot{A}}{A}-\frac{\dot{B}}{B}-\frac{\dot{C}}{C} = 0
\end{equation}
\begin{equation}
 \label{eq25}
\ddot{\phi}+\left(\frac{\dot{A}}{A}+\frac{\dot{B}}{B}+\frac{\dot{C}}{C}\right)
+\frac{r}{2\phi}{\dot{\phi}}^{2} = 0
\end{equation}
The energy conservation equation $T^{ij}_{;j} = 0$ yields
\begin{equation}
 \label{eq26}
\dot{\rho}+3(\rho+p)H-\lambda\frac{\dot{A}}{A} = 0
\end{equation}

\section{Solution of the field equations}
We have a system of six equations (\ref{eq20})$-$(\ref{eq26}) involving 
seven unknown variables namely, $A$, $B$, $C$, $p$, $\rho$, $\lambda$ and $\phi$. 
Therefore, in order to solve the field equations completely, we need at least one 
suitable physical assumption among the unknown variables. So, we constrain the 
system of equations with the law of variation for the Hubble's parameter proposed 
by Berman\cite{Berman1983}, which yields a constant value of DP. This law reads as 

\begin{equation}
 \label{eq27}
H = Da^{-n}
\end{equation}
where $D$ and $n$ are positive constants. 
In this paper, we show how the constant DP models with metric (\ref{eq1}) behave 
in the presence of string fluid and dimensionless scalar field $\phi$.\\
The deceleration parameter $(q)$, an important observational quantity, is defined as 
\begin{equation}
 \label{eq28}
q=-\frac{a\ddot{a}}{{\dot{a}}^{2}}
\end{equation}
From equations (\ref{eq4}) and (\ref{eq27}), we get
\begin{equation}
 \label{eq29}
\dot{a}=Da^{-n+1}
\end{equation}
Integration of (\ref{eq29}) leads to
\begin{equation}
 \label{eq30}
a=(nDt+c_{1})^{\frac{1}{n}},\;\;\;\;\; (n\neq 0)
\end{equation}
It is important to note here that for $n = 0$, the model has non 
singular origin and it evolves with exponential expansion which seems reasonable to project 
the dynamics of future Universe. Since we are looking for a model of Universe, 
which describe the dynamics of Universe from big bang to present epoch. 
Hence in this paper, the case $n = 0$ has been omitted.\\

Integrating equation (\ref{eq24}) and absorbing the constant of integration 
in $B$ or $C$, without loss of generality, we obtain
\begin{equation}
 \label{eq31}
A^{2} = BC
\end{equation}
Subtracting equation (\ref{eq21}) from equation (\ref{eq22}) and taking second 
integral, we get the following relation
\begin{equation}
 \label{eq32}
\frac{B}{C}=d_{1}exp\left[x_{1}\int{\frac{dt}{V}}\right]
\end{equation}
 where $d_{1}$ and $x_{1}$ are constants of integration.\\

From equations (\ref{eq3}), (\ref{eq30}), (\ref{eq31}) and (\ref{eq32}), the metric function 
can be explicitly written as
\begin{equation}
 \label{eq33}
A=(nDt+c_{1})^{\frac{1}{n}} 
\end{equation}
\begin{equation}
 \label{eq34}
B=\sqrt{d_{1}}(nDt+c_{1})^{\frac{1}{n}}exp\left[\frac{x_{1}}{2D(n-3)}(nDt+c_{1})^{\frac{n-3}{3}}\right]
\end{equation}
\begin{equation}
 \label{eq35}
C=\frac{1}{\sqrt{d_{1}}}(nDt+c_{1})^{\frac{1}{n}}exp\left[-\frac{x_{1}}{2D(n-3)}(nDt+c_{1})^{\frac{n-3}{3}}\right]
\end{equation}
provided that $n \neq 3$.\\

Inserting equation (\ref{eq4}) into equation (\ref{eq25}) and then integrating, we obtain
\begin{equation}
 \label{eq36}
\dot{\phi}^{2}\phi^{r} = d_{2}a^{-6}
\end{equation}
Here, $d_{2}$ is constant of integration.\\

The average's Hubble's parameter $(H)$, isotropic pressure $(p)$, proper energy density $(\rho)$, 
string tension density $(\lambda)$ and particle energy density $(\rho_{p})$ are found to be
\begin{equation}
 \label{eq37}
H=\frac{D}{nDt+c_{1}}
\end{equation}
\begin{equation}
 \label{eq38}
p=\alpha^{2}(nDt+c_{1})^{-\frac{2}{n}}-\left(\frac{2\omega d_{2}-x_{1}^{2}}{4}\right)(nDt+c_{1})^{-\frac{6}{n}} 
-3(1-n)D^{2}(nDt+c_{1})^{-2}
\end{equation}
\begin{equation}
 \label{eq39}
\rho=3D^{2}(nDt+c_{1})^{-2}-\frac{(x_{1}^{2}+2\omega d_{2})}{4}(nDt+c_{1})^{-\frac{6}{n}}
-3\alpha^{2}(nDt+c_{1})^{-\frac{2}{n}}
\end{equation}
\begin{equation}
 \label{eq40}
\lambda=\frac{(x_{1}^{2}-\omega d_{2})}{4}(nDt+c_{1})^{-\frac{6}{n}}
\end{equation}
\begin{equation}
 \label{eq41}
\rho_{p}=3D^{2}(nDt+c_{1})^{-2}-\frac{3x_{1}^{2}}{4}(nDt+c_{1})^{-\frac{6}{n}}-3\alpha^{2}(nDt+c_{1})^{-\frac{2}{n}}
\end{equation}
The above solutions satisfy the energy conservation equation (\ref{eq26}) identically, as expected.\\

The spatial volume $(V)$, expansion scalar $(\theta)$ and DP (q) are given by
\begin{equation}
 \label{eq42}
V = (nDt+c_{1})^{\frac{3}{n}}
\end{equation}
\begin{equation}
 \label{eq43}
\theta = 3D(nDt+c_{1})^{-1}
\end{equation}
\begin{equation}
 \label{eq44}
q=n-1
\end{equation}

We observe that at $t = -\frac{c_{1}}{nD}$, the spatial volume 
vanishes while all other parameters diverge. Therefore, the model has 
a big bang singularity at $t = -\frac{c_{1}}{nD}$. This singularity is point type because 
the directional scale factors $A(t)$, $B(t)$ and $C(t)$ vanish at the initial moment. 
From equation (\ref{eq43}), it is clear that for $n = 1$, the universe expands with 
constant rate. However, the recent observations of SN Ia (Perlmutter et al\cite{Perlmutter1997}$-$\cite{Perlmutter1999} 
Riess et al\cite{Riess1998,Riess2004} and Tonry et al\cite{Tonry2003}) reveal that the present 
Universe is accelerating and value of DP lies somewhere in the range $-1 < q < 0$. 
It follows that one can choose the value of $n$ in the range $0 < n < 1$ to have the 
consistancy of derived model with observations.\\ 

In the derived model, the scale factors increase with time. But the 
contribution of exponential terms to the scale 
factors $B$ and $C$ becomes negligible for sufficiently large time i. e. 
for sufficiently large time we have $A(t)\approx B(t)\approx C(t)$. 
This may be observed from equation (\ref{eq33})$-$(\ref{eq35}).    
Thus, initially the growth of scale factors take place at different rates 
due to effective contribution of exponential terms in $B$ and $C$. But later on 
the scale factors grow at the same rate. Therefore, in the derived model, the early 
anisotropic Universe becomes isotropic at later times.\\

The scalar function $(\phi)$ may be obtained as
\begin{equation}
 \label{eq45}
\phi=\left[\frac{\phi_{0}(r+2)}{2D(n-3)}(nDt+c_{1})^{\frac{n-3}{n}}\right]^{\frac{2}{r+2}}
\end{equation}
 where $\phi_{0}$ is the constant of integration.\\

The shear scalar $(\sigma)$ and anisotropy parameter $(A_{m})$ are read as
\begin{equation}
 \label{eq46}
\sigma = \frac{x_{1}}{2}(nDt+c_{1})^{-\frac{3}{n}}
\end{equation}
\begin{equation}
\label{eq47}
A_{m} = \frac{x_{1}^{2}}{6D^{2}}(nDt+c_{1})^{\frac{2n-6}{n}}
\end{equation}

\begin{figure}
\begin{center}
\includegraphics[width=4in]{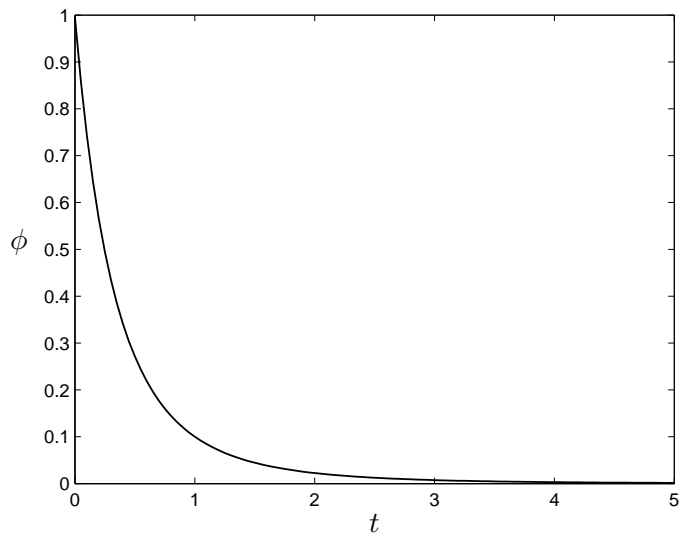} 
\caption{Plot of scalar function $(\phi)$ vs. time.} 
\label{fg:anil41F2.eps}
\end{center}
\end{figure}

\begin{figure}
\begin{center}
\includegraphics[width=4in]{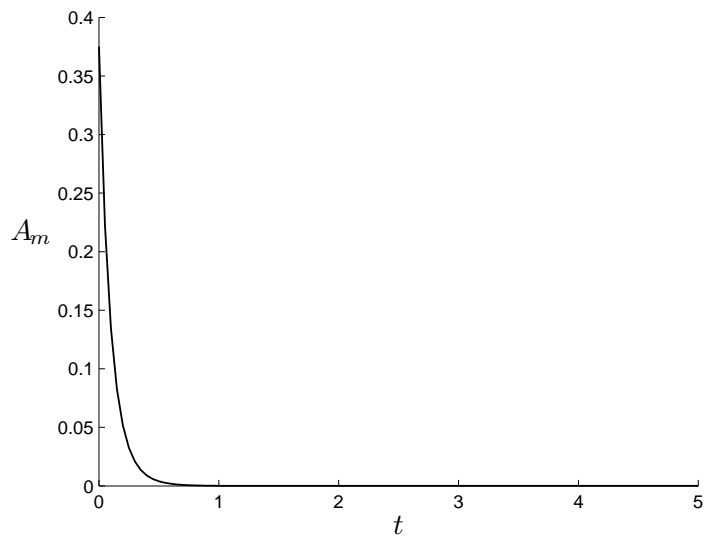} 
\caption{Plot of anisotropy parameter $(A_{m})$ vs. time }
\label{fg:anil41F3.eps}
\end{center}
\end{figure}

\begin{figure}
\begin{center}
\includegraphics[width=4in]{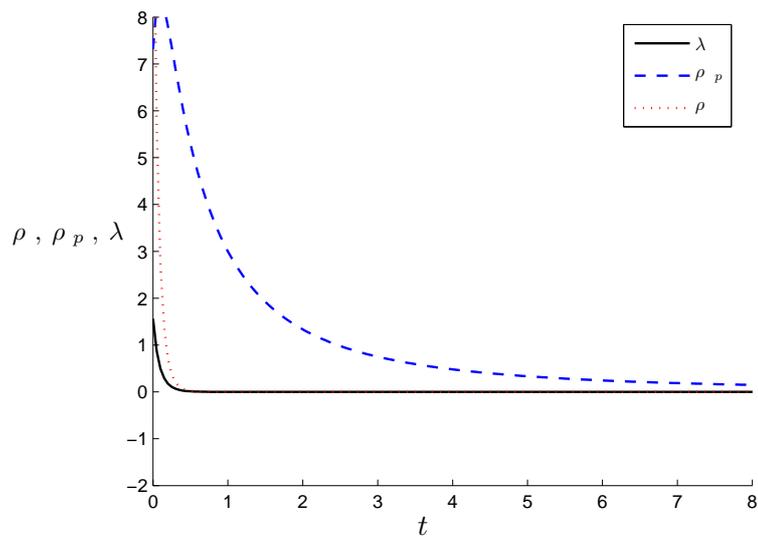} 
\caption{Plot of proper energy density $(\rho)$, string tension density $(\lambda)$ and 
particle energy density $(\rho_{p})$ vs. time.}
\label{fg:anil41F4.eps}
\end{center}
\end{figure}

\begin{figure}
\begin{center}
\includegraphics[width=4in]{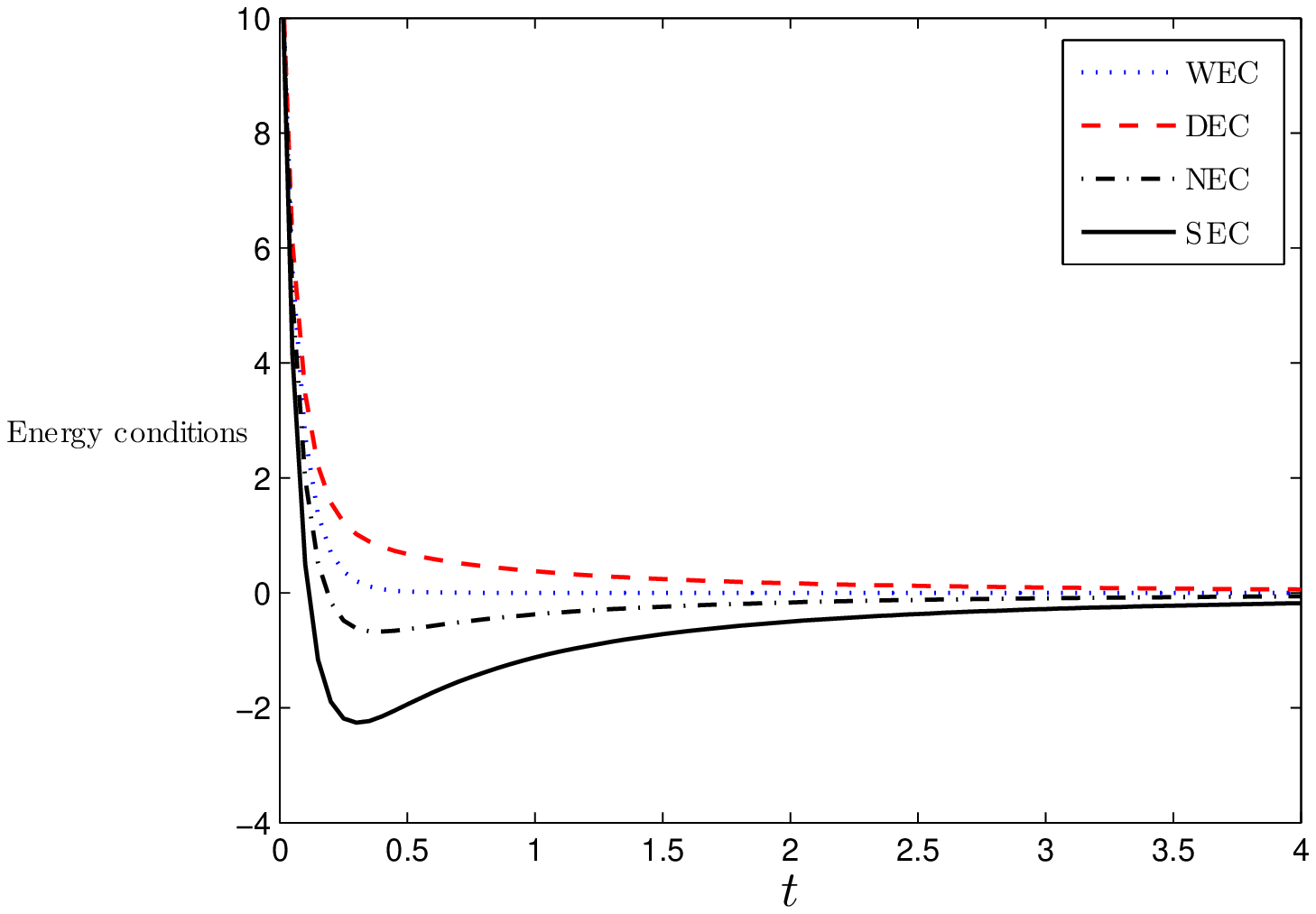} 
\caption{The left hand side of energy conditions vs. time.}
\label{fg:anil41F5.eps}
\end{center}
\end{figure}

The behaviours of $\rho$, $\rho_{p}$ and $\lambda$ are depicted in Figure 3. 
From eq. (\ref{eq40}) and (\ref{eq41}), it is clear that for $n < 1$ and for large value of 
time, $\frac{\rho_{p}}{\lambda} > 1$. This means that the particles dominate the strings at 
later times which confirms the disappearance of strings in the present day observations.\\

The behaviour of scalar function $(\phi)$ is depicted in figure 1. From equation (\ref{eq47}), 
it is clear that for $n < 1$, the anisotropy parameter $(A_{m})$ vanishes at late time. The behaviour 
of $A_{m}$ versus time is shown in figure 2.\\
\begin{figure}
\begin{center}
\includegraphics[width=4in]{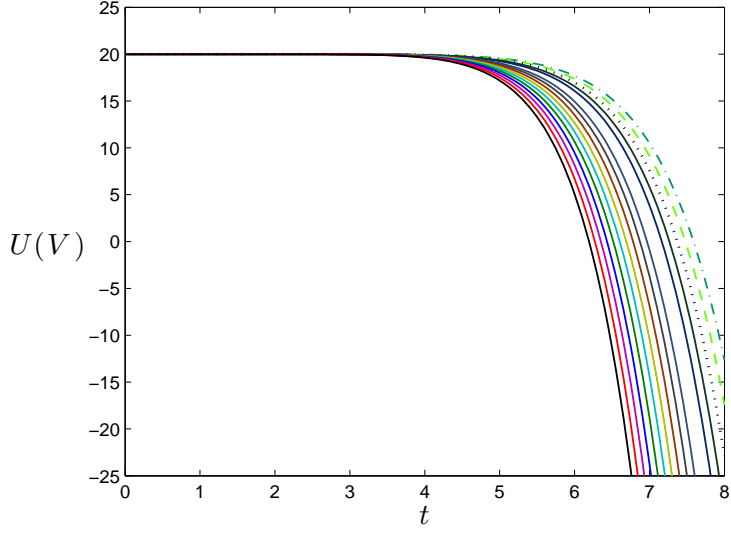} 
\caption{The classical potential vs. time.}
\label{fg:anil41F6.eps}
\end{center}
\end{figure}

From equation (\ref{eq42}), we obtain
\begin{equation}
 \label{eq48}
\dot{V} = 3D(nDt+c_{1})^{\frac{3-n}{n}}
\end{equation}
According to Saha and Boyadjiev \cite{Saha2004}, the equation of motion of a 
single particle with unit mass under force F(V) can be described as
\begin{equation}
 \label{eq49}
\dot{V} = \sqrt{2[\in - U(V)]}
\end{equation}
where $U(V)$ and $\in$ are the classical potential of force F and viewed energy level respectively.\\
From equations (\ref{eq48}) and (\ref{eq49}), we obtain
\begin{equation}
 \label{eq50}
U(V) = 2\in - 9D^{2}(nDt+c_{1})^{\frac{6-2n}{n}}
\end{equation}
In connection, with Hubble's parameter the classical potential $(U)$ is given by
\begin{equation}
 \label{eq51}
U(V) = 2\in - 9D^{\frac{6}{n}}H^{\frac{2n-6}{n}}
\end{equation}
Figure 4 plots the left hand side of energy conditions versus time. We observe that 
the weak energy condition (WEC) and dominant energy condition (DEC) are satisfied in 
the derived model. The null energy condition (NEC) is violated in the early universe but it 
is eventually satisfied in the present universe. It can also be observed that the strong energy 
condition (SEC) is violated in the derived model. The violation of SEC gives a reverse gravitation 
effect which may be possible cause for late time accelerated expansion of universe. 
Figure 5 plots the classical potential with respect to time in presence of string fluid as 
source of matter. We observe that $U(V)$ shows positive and negative nature with respect to time. \\  

\section{Conclusion}
In this paper, we have studied Bianchi - V string cosmological model in scalar - tensor 
theory of gravitation. The study reveals that the sting tension density $(\lambda)$ vanishes 
at present epoch that is why strings disappears from present universe but it was playing 
a significant role in the expansion of early universe. The derived model is singular in 
nature and it has big bang singularity at $t = -\frac{c_{1}}{nD}$. Thus the universe starts 
evolving from the infinite big bang singularity at $t = -\frac{c_{1}}{nD}$ and expands 
with power law expansion rate. The spatial volume is zero at initial moment $t = -\frac{c_{1}}{nD}$. 
At this instant, the physical parameters $\rho$, $p$, $\lambda$, $\rho_{p}$, $H$ and $\sigma$ all assume 
infinite values. These parameters are decreasing function of time and ultimately tend to zero 
for sufficiently large value of time. The spatial volume tends to zero as $t\rightarrow \infty$. 
Thus, the universe is essentially an empty space-time for large t.\\

The age of universe is given by
$$T=\frac{1}{(q+1)}H^{-1}-\frac{c_{1}}{(q+1)D}$$   
Thus the age of universe increases with $-1 < q < 0$ which shows the consistency of derived 
model with observations.\\

We have also discussed the classical potential with respect to time and have observed 
that the classical potential changes its nature with evolution of universe. In early universe, 
it is found positive and grows with constant rate but at late time, it is ruled with negative 
value and decreases rapidly with time.\\

\end{document}